\documentclass[prc,showpacs,preprintnumbers,amsmath,amssymb]{revtex4}
\usepackage{graphicx}
\newcommand{\beq}{\begin{equation}}
\newcommand{\eeq}{\end{equation}}
\newcommand{\beqa}{\begin{eqnarray}}
\newcommand{\eeqa}{\end{eqnarray}}
\newcommand{\nn}{\nonumber}
\newcommand{\Sigs}{\Sigma_{\mathrm s} }
\newcommand{\Sigv}{\Sigma_{\mathrm v} }
\newcommand{\Sigo}{\Sigma_{\mathrm o} }

\newcommand{\bfgamma}{\mbox{\boldmath$\gamma$\unboldmath}}

\newcommand{\Sigsi}{\Sigma_{{\mathrm s},i} }

\newcommand{\Sigoi}{\Sigma_{{\mathrm o},i} }
\begin{document}
\preprint{}
\title{Spinodal Instabilities in Asymmetric Nuclear Matter Based on Realistic $NN$ Interactions.}
\author{E. N. E. van Dalen}
\author{H. M\"uther}
\affiliation{Institut f$\ddot{\textrm{u}}$r Theoretische Physik,
Universit$\ddot{\textrm{a}}$t T$\ddot{\textrm{u}}$bingen, Auf der
Morgenstelle 14, D-72076 T$\ddot{\textrm{u}}$bingen, Germany}
\begin{abstract}
A density dependent relativistic mean-field model is determined to reproduce the
components of the nucleon self-energy at low densities. This  model is used to 
investigate spinodal instabilities in isospin asymmetric nuclear matter at
finite temperatures. The inhomogeneous density distributions in the spinodal
region are investigated through calculations in a cubic Wigner-Seitz cell.
Compared to results obtained in phenomenological calculations  the spinodal
region  is large, i.e. the spinodal region at zero temperature can reach
densities above 0.12 fm$^{-3}$.  The predicted spinodal region is 
concentrated around isospin symmetric nuclear matter and the critical
temperature is considerably lower than in the previous microscopic based
investigation within a non-relativistic Brueckner-Hartree-Fock approach. 
\end{abstract}
\maketitle
\section{Introduction}
The properties of isospin asymmetric nuclear matter systems like their
instabilities and phase transitions are presently a topic  of great
interest~\cite{margueron:2003,chomaz:2004,avancini:2004,avancini:2006,margueron:2007,vidana:2008,pais:2009}.
At either high energy or high baryon density one explores the phase transition
to quark matter, i.e. the hadrons  dissolve into a phase of de-confined quarks.
Below a critical temperature or density, the homogeneous fluid consisting of
nucleons may condense into nuclei. This transition  can be regarded to be of the
liquid-gas type. Heavy ion collisions are well suited to investigate these
transitions, since the multi-fragmentation phenomenon occurs in these
experiments. In the multi-fragmentation phenomenon, highly excited composed
nuclei are formed in a gas of evaporated particles. Therefore, this
multi-fragmentation phenomenon can be interpreted as the coexistence of a liquid
and a gas phase. This liquid-gas phase transition is important in the
description of the nucleation processes. Furthermore, it plays an important role
in the collapse of supernovae into neutron stars.

The hadrons involved in this transition can be either neutrons or protons.
Therefore, one should consider nuclear matter as composed of two different
fluids. These phase transitions in nuclear matter are related to the
thermodynamic instability regions, which are limited by so-called spinodals.
This spinodal instability is dominated by mechanical or isoscalar density
fluctuations which lead to a liquid-gas phase separation with restoration of the
isospin symmetry in the dense phase. This phenomenon is known as the isospin
distillation or fragmentation effect~\cite{xu:2000}, i.e. the formation of 
large droplets of high density symmetric nuclear matter within a neutron gas
with only a very small fraction of protons. 

In the past systematic analysis of the stability conditions of asymmetric
nuclear matter against the liquid-gas phase transition have been performed using
different methods. These approaches can roughly be divided into phenomenological
methods and microscopic based methods. Phenomenological studies of spinodal
instabilities involve mean-field calculations with effective forces of the
Skyrme or Gogny type~\cite{margueron:2003,ducoin:2006,ducoin:2007a,ducoin:2007b}
and relativistic mean-field calculations using constant and density dependent
coupling
parameters~\cite{liu:2002,avancini:2004,providencia:2006,avancini:2006,pais:2009}.
Some of these models predict that the total density at which spinodal
instability appears will decrease if the asymmetry increases, whereas for other
models it will increase up to large asymmetry and will finally decrease.
However, a common feature of these models is that the spinodal region is in
principle situated at densities below $\rho_B < 0.1$ fm$^{-3}$. The microscopic
based studies are a calculation with  low-density
functionals~\cite{margueron:2007}  based on the Dirac-Brueckner-Hartree-Fock
(DBHF) approach~\cite{vandalen:2004b} with a realistic Bonn A
potential~\cite{machleidt:1987}  and a calculation within a microscopic
Brueckner-Hartree-Fock (BHF) approach~\cite{vidana:2008} using the realistic
Argonne V18 potential~\cite{wiringa:1995} plus a three-body force of Urbana
type.  Both calculations predict larger spinodal regions compared to those
obtained in the phenomenological calculations, i.e. the spinodal region can
reach densities above 0.12 fm$^{-3}$ close to symmetric nuclear matter. The
critical temperature in the BHF calculation is $T_c=17.5$ MeV, whereas finite
temperature effects are not considered in the work of
Ref.~\cite{margueron:2007}.

In the present work, we construct a density dependent relativistic mean-field
(DDRMF) model, i.e. a density dependent relativistic Hartree (DDRH) model, which
is based on microscopic DBHF calculations~\cite{vandalen:2007} using the
realistic Bonn A potential~\cite{machleidt:1987}. Apart from the application  of
the results of this DBHF calculation to spinodal
instabilities~\cite{margueron:2007}, it has also already been used for other
applications such as  investigations of the non-homogeneous structures in the
neutron star crust~\cite{gogelein:2008}, investigations of hybrid
stars~\cite{klahn:2007}, and  studies of nucleon-nucleus
scattering~\cite{xu:2012}. In the previous DBHF based investigation of the
spinodal instability~\cite{margueron:2007}, a functional is constructed on a
density expansion of the relativistic mean-field theory with fixed coupling
constants. However, in the present work the approximation of an expansion is not
made and the effective coupling functions, which are obtained from the DBHF
self-energy components, are density dependent. Special attention is devoted to
reproduce the DBHF self-energy also at very low densities, i.e. $\rho_B < 0.10
fm^{-3}$. In the framework of this DDRH theory based on a DBHF approach, the
equation of state at zero and finite temperature can then be determined from an
iteration procedure until convergence is reached. Within this model, the
temperature effects and the nature of spinodal instabilities are investigated.
In addition to these investigations in infinite nuclear matter calculations, the
density distribution in the spinodal region is studied in cubic Wigner-Seitz
(WS) cell calculations.

The paper is organized as follows. The density dependent relativistic Hartree
(DDRH) theory is treated in Sec.~\ref{sec:DDRHF}. The parameterization for the
DDRH theory based on the DBHF results are discussed in
Sec.~\ref{sec:parametrization}. Section~\ref{sec:si} is devoted to the theory of
the spinodal instabilities. The results for the spinodal instabilities  are
presented and discussed in Sec.~\ref{sec:R}. Finally, we end with a summary and
the conclusion in Sec.~\ref{sec:S&C}.

\section{Density Dependent Relativistic Hartree Theory.}
\label{sec:DDRHF}

A Lagrangian density of an interacting many-particle
system consisting of nucleons and mesons is the starting
point of a DDRH theory. The Lagrangian density of the DDRH theory used for our investigation includes as well the isoscalar mesons $\sigma$ and $\omega$ as the isovector mesons $\delta$ and $\rho$. Therefore, the Lagrangian density has the following parts:
the free baryon Lagrangian density $\mathcal{L}_B$,
the free meson Lagrangian density $\mathcal{L}_M$, and
the interaction Lagrangian density $\mathcal{L}_{\text{int}}$:
\begin{equation}\label{Lag_dens}
	\mathcal{L} = \mathcal{L}_B + \mathcal{L}_M + \mathcal{L}_{\text{int}},
\end{equation}
which can be written as
\begin{equation}
\mathcal{L}_B =   \bar{\Psi} ( \, i \gamma _\mu \partial^\mu - M ) \Psi,  
\end{equation}
\begin{equation}
  \mathcal{L}_M = {\textstyle \frac{1}{2}} \sum_{\iota= \sigma, \delta}
			\Big( \partial_\mu \Phi_\iota \partial^\mu \Phi_\iota - m_\iota^2 \Phi_\iota^2 \Big)   
  		 	- {\textstyle \frac{1}{2}} \sum_{\kappa = \omega, \rho}
			\Big( \textstyle{ \frac{1}{2}} F_{(\kappa) \mu \nu}\, F_{(\kappa)}^{\mu \nu}
				- m_\kappa^2 A_{(\kappa)\mu} A_{(\kappa)}^{\mu} \Big),      
\end{equation}
\begin{equation}
  \mathcal{L}_{\text{int}} =	- g_\sigma\bar{\Psi}  \Phi_\sigma \Psi
                - g_\delta \bar{\Psi}  \boldsymbol{\tau} \boldsymbol{\Phi}_\delta \Psi
		 - g_\omega \bar{\Psi}  \gamma_\mu A_{(\omega)}^{ \mu } \Psi  
		- g_\rho \bar{\Psi}  \boldsymbol{\tau } \gamma_\mu  \boldsymbol{A}_{(\rho)}^{\mu } \Psi. 
\label{eq:Lint}
\end{equation}
In the above Lagrangian density the field strength tensor for the vector mesons is defined as
$F_{(\kappa)\mu \nu} = \partial_{\mu} A_{(\kappa)\nu} - \partial_\nu A_{(\kappa)\mu}$. 
The nucleon field is  $\Psi$
and the nucleon rest mass is $M$.
The meson fields are denoted by $ \Phi_\sigma$,$ \boldsymbol{\Phi}_\delta $, $ A_{(\omega)} $, and $ \boldsymbol{A}_{(\rho)} $. Moreover, the bold symbols denote vectors in the isospin space.
The mesons couple to the nucleons with the strength of the coupling constants
$ g_\sigma$, $g_\delta$, $g_\omega$, and $g_\rho$. The rest masses of these mesons are $m_\sigma$, $m_\omega$, $m_\delta$, and $m_\rho$. 

By minimizing the action
for variations of the fields $\bar{\Psi}$
included in the Lagrangian density of Eq.~(\ref{Lag_dens}), one obtains the field equations for the nucleons, 
\begin{equation}
  \delta \int_{t_0}^{t_1} dt \int d^3x \,
      \mathcal{L}\big(\bar{\Psi}(x), \partial_\mu \bar{\Psi}(x), t \big) = 0.
\end{equation}
Finally the following field equation can be derived 
\begin{equation}
\frac{\partial}{\partial x^\mu}
\frac{\partial \mathcal{L} }{\partial(\partial_\mu \bar{\Psi})}
-  \frac{\partial \mathcal{L}}{\partial \bar{\Psi}}
	- \frac{\partial \mathcal{L}}{\partial \rho} \frac{\delta \rho}{\delta \bar{\Psi}}= 0.
\label{eq:ELE}
\end{equation}
Without density dependent meson-baryon vertices the third term vanishes in Eq.~(\ref{eq:ELE}) and 
the normal Euler-Lagrange field equation is obtained. From the evaluation of Eq.~(\ref{eq:ELE}),
we obtain the Dirac equation,
\begin{equation}
\big(i\gamma_\mu \partial^\mu - M - ( \Sigma + \Sigma^{(r)} \gamma_0 )\big) \, \Psi = 0,
\label{eq:Dirac}
\end{equation}
with  $\Sigma$ the nucleon self-energy and $\Sigma^{(r)}$ the so-called rearrangement
contribution to the nucleon self-energy. 

The self-energy $\Sigma$ for the Lagrangian density in Eqs.~(\ref{Lag_dens})-(\ref{eq:Lint}) can be written as 
\begin{equation}
\begin{split}
  \Sigma =   g_\sigma \Phi_\sigma 
		+ g_\delta \boldsymbol{\tau} \boldsymbol{\Phi}_\delta
		+ g_\omega \gamma_\mu A_{(\omega)}^{ \mu }   
	        + g_\rho \boldsymbol{\tau } \gamma_\mu  \boldsymbol{A}_{(\rho)}^{\mu }.
\end{split}
\label{eq:sigma_1}
\end{equation}
behavior under Lorentz transformations,  
\beqa
\Sigma(k)= \Sigs (k) -\gamma_0 \, \Sigo (k) + 
\bfgamma  \cdot \textbf{k} \,\Sigv (k).
\label{subsec:SM;eq:self1}
\eeqa
The $\Sigs$, $\Sigo$, and $\Sigv$ components are Lorentz scalar
functions. Hence, one can define the following effective quantities
\begin{equation}
k^*=k(1+\Re e \Sigma_v(k)), 
\end{equation}
\begin{equation}
m^*(k) = M + \Re e \Sigma_s(k), 
\end{equation}
and
\begin{equation}
E^*(k)=E(k) + \Re e \Sigma_o(k).
\end{equation}
In the on-shell case, the effective energy can also be given by
\begin{equation}
E^*(k)^2={k^*}^2+m^*(k)^2.
\end{equation} 
The contributions  to the self-energy of the Lagrangian density presented in Eqs.~(\ref{Lag_dens})-(\ref{eq:Lint}) can be written as
\beqa
\Sigsi (k) & = & - \left(\frac{g_{\sigma}}{m_{\sigma}}\right)^2 (\rho_{s,n} + \rho_{s,p}) - \left(\frac{g_{\delta}}{m_{\delta}}\right)^2
\sum_{j=n,p} (\rho_{s,i}-\rho_{s,j})  
\nn 
\eeqa
and
\beqa
\Sigoi (k) & = & -\left(\frac{g_{\omega}}{m_{\omega}}\right)^2 (\rho_n + \rho_p) - \left(\frac{g_{\rho}}{m_{\rho}}\right)^2 \sum_{j=n,p} (\rho_{i}-\rho_{j}),  \nn 
\eeqa
\beqa
  \rho_{s,i} &=& \frac{1}{\pi^2} \int f_i(k) k^2  dk   \frac{m_i^\ast}{E_i(k)^\ast}  
\label{eq:NM_sdens} 
\eeqa
and
\beqa
\rho_i &=& \frac{1}{\pi^2} \int f_i(k) k^2  dk.
\label{eq:NM_dens}
\eeqa
The thermal occupation factors in Eqs.~(\ref{eq:NM_sdens}) and (\ref{eq:NM_dens}) are given by 
\begin{equation}
  f_i (k) = \frac{1}{1+ \exp[ (E_i^\ast - \varepsilon_{F,i}) / T ]}, \ \ \ i = p,n
\end{equation}
in the ''no-sea'' approximation.

The $\Sigma^{(r)}$ term in Eq.~(\ref{eq:Dirac}) is generated by the third term in Eq.~(\ref{eq:ELE}). It will only give a nonzero contribution, if meson-baryon vertices are density dependent. This rearrangement contribution $\Sigma^{(r)}$ can be written as
\begin{equation}\label{self_en_rearr_1}
\begin{split}
  \Sigma^{(r)} = \Big(
	& \frac{\partial g_\sigma}{\partial \rho} \bar{\Psi} \Phi_\sigma \Psi
	 + \frac{\partial g_\delta}{\partial \rho} \bar{\Psi} \boldsymbol{\tau} \boldsymbol{\Phi}_\delta \Psi\\ &
	 + \frac{\partial g_\omega}{\partial \rho} \bar{\Psi} \gamma_\mu A_{(\omega)}^{ \mu } \Psi 
	  + \frac{\partial g_\rho}{\partial \rho}
	      \bar{\Psi} \boldsymbol{\tau } \gamma_\mu  \boldsymbol{A}_{(\rho)}^{\mu } \Psi
            \Big).
\end{split}
\end{equation}
Rearrangement contributions are essential
to provide a symmetry conserving approach~\cite{fuchs:1995,vandalen:2010b}, i.e. an approach satisfying energy-momentum 
conservation and thermodynamic consistency like the Hugenholtz - van Hove
theorem.

\section{Parameterization}
\label{sec:parametrization}
We have constructed a density dependent relativistic mean-field (DDRMF) model,
i.e. a density dependent relativistic Hartree (DDRH) model.  The effective
density dependent coupling functions of this DDRH model are determined by the
requirement that the results of a DBHF calculation at zero temperature using Bonn
A~\cite{vandalen:2007} for the components of the nucleon self-energy at the
Fermi momentum are reproduced. The rearrangement terms of DDRH are not included in the
fit, since the DBHF approach has no rearrangement contributions. 
\begin{table}
\begin{center}
\begin{tabular}{|c|c|ccccc|}
\hline
\ \ \ meson i & \ $m$ [MeV]\ \ \ &\ \ \ $a_i$\ \ \ &\ \ \ $b_i$ \ \ \ 
& \ \ \ $c_i$  \ & \ \ \ $d_i$ \ \ \  & \ \ \ $e_i$ \ \ \ 	\\
\hline
 $\sigma$ & 550 & $14.0089064$     & $9.91395378 $    & $-2.1599853$   &  $0.280160069$ &   $-13.3718958$ \\
 $\omega$ & 782.6 & $13.5288515 $    & $7.86182976$    & $-1.8871752$   & $0.244318038$  &  $-9.42372704$ \\
 $\rho $ & 769  & $4.90660906$   & $-14.9182243$    & $3.93277073$    & $-0.498567462$ &  $13.1945915$ \\
 $\delta$   & 983    & $-8.32733536$    & $-43.6584053$    & $10.4067049$   & $-1.26465869$ &   $50.3025665$    \\
\hline
\end{tabular}
\end{center}
\caption{\label{table:DDRH_parameters_swrd} Parameter set for the DDRH  model from the DBHF approach in 
Ref.~\cite{vandalen:2007}.}
\end{table}
     
The density dependence of the coupling functions are parameterized by
\begin{equation}
  g_i (\rho _B)
    = a_i + b_i x + c_i  x^2 + d_i x^3 + e_i \sqrt{x},
\label{eq:EvD_coupl_func}
\end{equation}
with $x = \rho _B/\rho_0$ and $\rho_0$ = 0.16 fm$^{-3}$. The DBHF data at densities of $\rho_B$ = 0.017, 0.100, 0.197, 0.313, and
$0.467$ fm$^{-3}$ at a proton fraction of  $Y_p = 0.4$ are used to obtain these parameters of our DDRH model. The meson masses 
are chosen to be identical to those of the Bonn A potential. All parameters are given in table \ref{table:DDRH_parameters_swrd}.
\begin{figure}[!t]
\begin{center}
\includegraphics[width=0.6\textwidth] {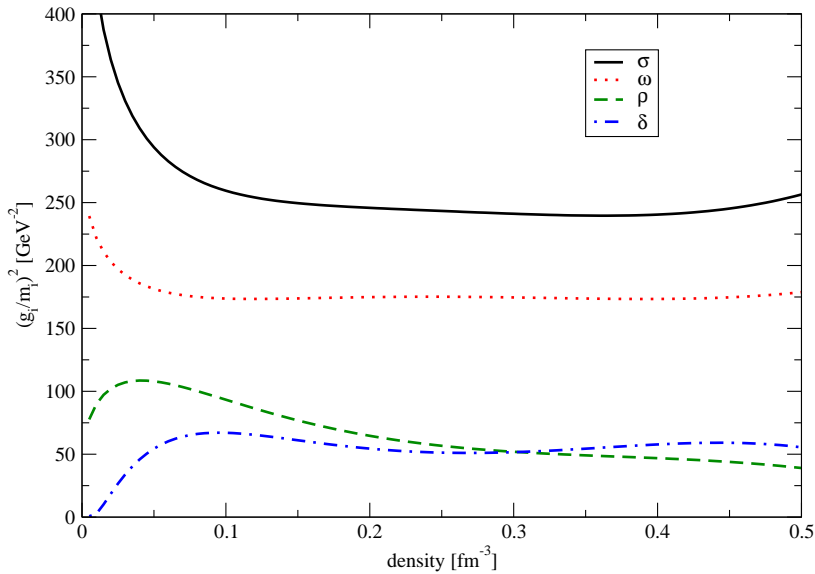} 
\caption{(Color online) Density dependence of the coupling functions of the $\sigma$-meson (solid), the $\omega$-meson (dotted), the $\rho$-meson (dashed), and the $\delta$-meson (dashed-dotted).}
\label{fig:coupling}
\end{center}
\end{figure}
The density dependence of these coupling functions is displayed in
Fig.~\ref{fig:coupling}. The density dependence is rather weak except at very
low densities. The strength for the isovector mesons ($\rho$ and $\delta$) is
significantly smaller than for the isoscalar mesons ($\sigma$ and $\omega$).
All coupling functions tend to decrease with increasing density. This may
reflect the quenching of the DBHF $G$-matrix underlying this fit due to Pauli
and dispersive effects in the Bethe-Goldstone equation. 

\begin{figure}[!t]
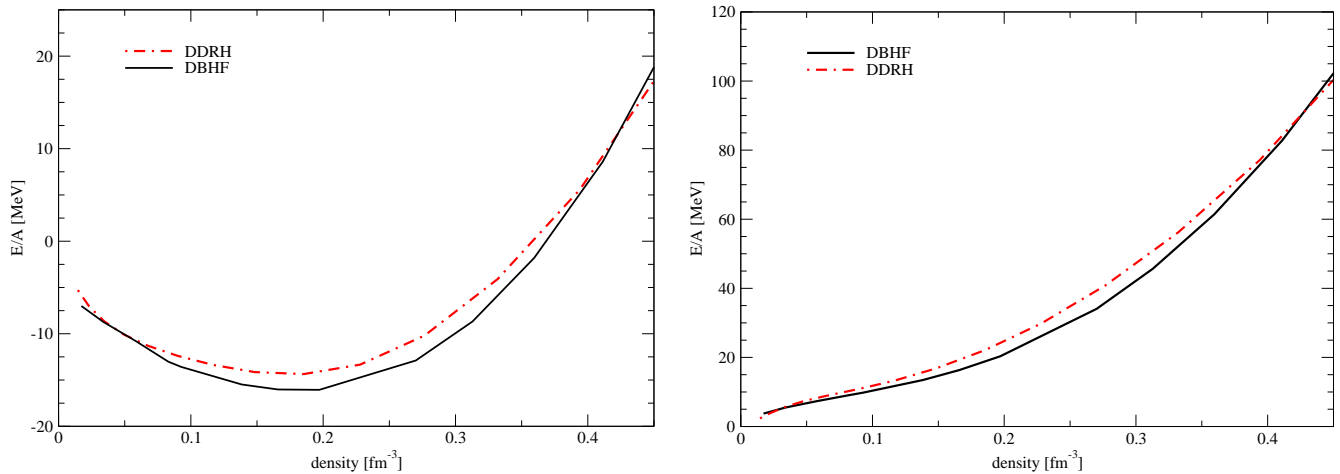

\begin{center}
\includegraphics[width=0.48\textwidth] {bindinglow.eps} \quad
\includegraphics[width=0.48\textwidth] {neutmatterlow.eps}
\caption{(Color online) Binding energy is plotted as a function of the density for the DBHF approach and the DDRH model. Left: isospin symmetric nuclear matter. Right: pure neutron matter.}
\label{fig:binding}
\end{center}
\end{figure}
\begin{table}
\begin{center}
\begin{tabular}{|ll|c|c|}
\hline
&& DDRH & DBHF\\

\hline	  
$\rho_{sat}$ &       [fm$^{-3}]$ & 0.1759 \ &  0.181 \\
$E/A$   &[MeV]	            & -14.39  \ & -16.15 \\
$K$     &[MeV]              & 193    \  & 230 \\
$a_s$ & [MeV]        & 35.03   \  & 34.16 \\
\hline
\end{tabular}
\end{center}
\caption{\label{table:nm_prop}
Saturation properties of nuclear matter for the DDRH model compared to the DBHF results
of Ref.~\cite{vandalen:2004b,vandalen:2007}. The quantities listed include the saturation density $\rho_{sat}$, the
binding energy $E/A$, the compressibility modulus $K$, and the symmetry energy $a_s$. The symmetry energy $a_s$
has been obtained assuming a quadratic
dependence of the EOS on the asymmetry parameter $\beta=(\rho_n-\rho_p)/(\rho_n + \rho_p)$.}
\end{table}

In Fig.~\ref{fig:binding} the binding energy of the DDRH model is shown for
neutron matter and symmetric nuclear matter. The results are in reasonable
agreement with the DBHF results. Note, that the DDRH parameterization has solely
been based on the DBHF self-energies at the corresponding Fermi momenta for a
proton fraction $Y_p = 0.4$. No attempt has been made to account for the
momentum and further asymmetry dependence of the DBHF self-energies, which
enter the calculated energies for symmetric nuclear matter and neutron matter.
Both approaches show a nonlinear convergence to
zero when the density decreases.  which is in qualitative agreement with a study
of low-density nuclear matter~\cite{horowitz:2006} based on a virial expansion
that includes protons, neutrons, and $\alpha$-particle degrees of freedom. In
particular at densities below $\rho < 0.1 \textrm{fm}^{-3}$, the DDRH
model presented here is in better agreement with the DBHF result than those
models constructed in Ref.~\cite{vandalen:2011}. 

From table~\ref{table:nm_prop}, one can see that the saturation density $\rho_{sat}$  is shifted
to lower densities and the binding energy $E/A$  of the saturation point is weaker in the DDRH model compared to the original DBHF results.
The compression modulus $K$ is somewhat smaller and symmetry energy $a_s$ is in a fairly good agreement with the DBHF results.

\section{Spinodal Instabilities} 
\label{sec:si}
The stability conditions for isospin asymmetric nuclear matter at finite temperatures are obtained from the free energy density $\mathcal{F}$.
This free energy density is defined as 
\begin{equation}
  \mathcal{F} = \mathcal{E} - T \mathcal{S}, 
\end{equation}
where $\mathcal{E}$ is the energy density and $\mathcal{S}$ denotes the entropy density.
The energy density $\mathcal{E}$ is obtained from the energy-momentum tensor
\begin{eqnarray}
\mathcal{E} = \langle T^{00} \rangle
	&=& \frac{1}{\pi^2} \sum_{i=p,n} \int f_i(k)\, k^2 \, dk \,
		\sqrt{ k^2 + m_i^{\ast^2} }		\notag		\\
	&& + \frac{1}{2} \sum_{i=p,n} \left( \Sigsi \, \rho_{s,i} + \Sigoi  \, \rho_i \right), 
\label{eq:energyden}
\end{eqnarray}
where $f_i(k)$ is the thermal occupation probability. 
The rearrangement term $\Sigma^{(r)}$ does not appear in Eq.~(\ref{eq:energyden}), since rearrangement does not contribute to the energy density.
However, note that the rearrangement does modify other quantities such as the pressure.
Furthermore, the entropy density $\mathcal{S}$ is given by 
\begin{equation}
  \mathcal{S} = - \frac{1}{\pi^2} \sum_{i=p,n} \int f_i(k)\, k^2 \, dk 
		  \left[ f_i(k) \, \ln(f_i(k)) + (1-f_i(k) ) \, \ln(1-f_i(k)) \right].
\end{equation}

The necessary ingredients to analyze the stability of isospin asymmetric nuclear matter against 
separation into two phases are now introduced. The nuclear system will be stable if the free energy of a single phase is lower than the
free energy in any two-phase configurations. This  stability condition  implies that the free
energy density is a convex function of the neutron and proton densities $\rho_n$ and $\rho_p$.
This means that the curvature matrix, 
\begin{equation}
\left[ \mathcal{F}_{ij}\right] =\left[ \frac{\partial ^{2}\mathcal{F}}{
\partial \rho _{i}\partial \rho _{j}} \right]_T,  \label{eq:curvature}
\end{equation}
is positive-definite. The requirement that the local curvature is positive 
demands that both the trace and the determinant are positive,
\begin{equation}
\mathrm{Tr}[\mathcal{F}_{ij}]\geq 0,\quad \textrm{and} \quad \mathrm{Det}[\mathcal{F}_{ij}]\geq 0.  \label{eq:condition}
\end{equation}
The $[\mathcal{F}_{ij}]$ is a $2*2$ symmetric matrix, since one considers a two-fluids
system. Therefore, the stability matrix $[\mathcal{F}_{ij}]$ has two
real eigenvalues $\lambda ^{\pm}$~: 
\begin{equation}
\lambda ^{\pm}=\frac{1}{2}\left( \mathrm{Tr}\left[ \mathcal{F}_{ij}\right]
\pm \sqrt{\mathrm{Tr}\left[ \mathcal{F}_{ij}\right] ^{2}-4\mathrm{Det}\left[ 
\mathcal{F}_{ij}\right] }\right)  \label{eq:eigenvalues}
\end{equation}
associated to 
eigenvectors $\mathbf{\delta \rho }^{\pm}$ given by ($i\neq j$) 
\begin{equation}
\frac{{\delta \rho }_{j}^{\pm}}{{\delta \rho }_{i}^{\pm}}=\frac{\mathcal{F}
_{ij}}{\lambda ^{\pm}-\mathcal{F}_{jj}}=\frac{\lambda ^{\pm}-\mathcal{F}_{ii}
}{\mathcal{F}_{ij}}.  \label{eq:eigenvectors}
\end{equation}
Therefore, the requirements in Eq.~(\ref{eq:condition}) can equivalently be written as
\begin{equation}
\mathrm{Tr}[\mathcal{F}_{ij}] = \lambda ^{+}+\lambda ^{-} \geq 0 \quad \textrm{and} \quad \mathrm{Det}[\mathcal{F}_{ij}] = \lambda ^{+}\lambda ^{-} \geq 0.  \label{eq:condition2}
\end{equation}
This means that both eigenvalues should be positive to guarantee the stability of the system. Due to the essentially quadratic dependence of the symmetry energy on the asymmetry parameter and a positive increasing symmetry energy with the total density, it turns out that only one eigenvalue 
can become negative~\cite{margueron:2003}. In principle $\lambda^+$ will always be positive in isospin asymmetric nuclear. Only $\lambda^-$ can  become negative and the eigenvector associated with this negative eigenvalue indicates the
direction of the instability. Therefore, if Eq.~(\ref{eq:condition2}) is violated, or equivalently $\lambda^-$ is negative for isospin asymmetric nuclear matter, the system will be in the unstable region of a phase transition.

\section{Results and Discussion}
\label{sec:R}
\begin{figure}[!t]
\begin{center}
\includegraphics[width=0.6\textwidth]{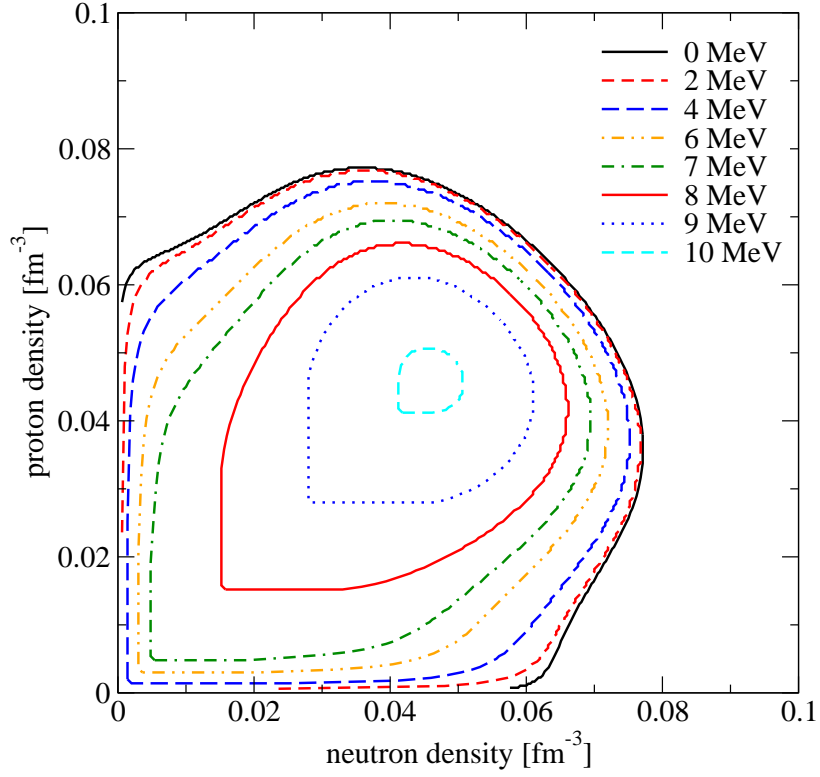} 
\caption{(Color online) The projection of the spinodal contour in the neutron-proton density plane at several temperatures. } 
\label{fig:spinodal}
\end{center}
\end{figure}

In the present work, we have investigated the spinodal region in the case of
isospin asymmetric nuclear matter at finite temperature. The projection of the
spinodal contour in the neutron-proton density plane for several temperatures
ranging from T=0 to T=10 MeV  is depicted in Fig.~\ref{fig:spinodal}. Keep in
mind that the isospin symmetry of the nucleon-nucleon interaction guarantees
the symmetry of the spinodal contours with respect to the interchange of the
neutron density $\rho_n$ and the proton density $\rho_p$. 

Let us first consider the results for zero temperature (T= 0 MeV).  One finds
that the spinodal region extends up to a maximum density of 0.127 fm$^{-3}$. 
This is considerably larger than the maximum density obtained in  phenomenological
models~\cite{margueron:2003,chomaz:2004,avancini:2004,avancini:2006,ducoin:2006,ducoin:2007a,ducoin:2007b,liu:2002,providencia:2006,pais:2009},
which indicate that the spinodal region is situated at densities
below $\rho_B < 0.1$ fm$^{-3}$. The large spinodal region, however, is in agreement
with microscopic calculations~\cite{margueron:2007,vidana:2008}. This difference
between the phenomenological and the microscopic based models is of course related
to the  larger saturation densities, which are obtained in the microscopic
calculations. Note, however, that the present study yields a larger maximal
density than the non-relativistic BHF calculations of Vidana and
Polls~\cite{vidana:2008}, although the saturation density for the present
approach ($\rho_0=0.176$ fm$^{-3}$) is smaller than the one obtained in the BHF
calculation ($\rho_0=0.182$ fm$^{-3}$).  

The shape of the spinodal region for T=0 projected in the neutron-proton density plane,
is similar to the shapes obtained in calculations before. In detail, however, there is a
difference: The region of instability is more confined to isospin symmetric matter.

Also the dependence of the region of instability on temperature in
Fig.~\ref{fig:spinodal} is very similar to the one observed in previous
calculations~\cite{margueron:2003,chomaz:2004,avancini:2004,avancini:2006,ducoin:2006,ducoin:2007a,ducoin:2007b,liu:2002,providencia:2006,vidana:2008,pais:2009}.
The spinodal region  shrinks with increasing temperature, until the critical temperature
$T_c$ is reached. In the present approach no spinodal region can be obtained above $T=11$
MeV. Therefore, the critical temperature is considerably lower than the value $T_c=17.5$
MeV obtained in the other microscopic based calculation~\cite{vidana:2008} or the values
for $T_c$ reported in the phenomenological studies~\cite{avancini:2006}. It is worth
mentioning that the spinodal region obtained here at temperatures close to $T_c$ 
occurs at slightly higher densities than in the other studies.

\begin{figure}[!t]
\begin{center}
\includegraphics[width=0.6\textwidth]{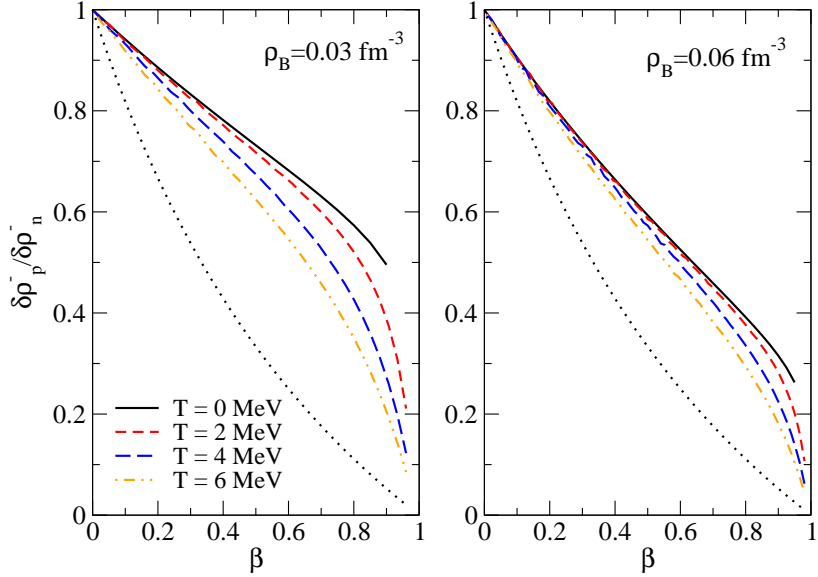} 
\caption{(Color online) Ratio $\delta \rho^{-}_p/\delta \rho^{-}_n$
as a function of asymmetry parameter $\beta=(\rho_n-\rho_p)/(\rho_n + \rho_p)$ for several temperatures. The left panel shows the results for a fixed density of $\rho_B = 0.03$ fm$^{-3}$, whereas in the right panel the results for a fixed density of  $\rho_B = 0.06$ fm$^{-3}$ are presented. The ratio $\rho_p/\rho_n$ is given by the black dotted line. 
 \label{fig:ratio}}
\end{center}
\end{figure}
It has been shown that the spinodal instabilities in isospin asymmetric nuclear appear as
a mixture of isoscalar and isovector instabilities~\cite{margueron:2003,chomaz:2004}.
Isoscalar or mechanical instabilities are related to density fluctuations. Pure isoscalar
instabilities conserve the proton fraction and fulfill the relation $\delta
\rho^{-}_p/\delta \rho^{-}_n = 1$. On the other hand, pure isovector or chemical
instabilities are related to fluctuations in the proton fraction at fixed density and
fulfil the relation $\delta \rho^{-}_p/\delta \rho^{-}_n = -1$.  Therefore, the
instabilities will be predominantly of the isoscalar nature, if the neutrons and protons
move in phase $\delta \rho^{-}_p/\delta \rho^{-}_n > 0$, or of the isovector nature, if
they move out of phase $\delta \rho^{-}_p/\delta \rho^{-}_n < 0$. In Fig.~\ref{fig:ratio}
the results obtained from our model for this ratio $\delta \rho^{-}_p/\delta \rho^{-}_n$ 
is plotted as a function of the asymmetry parameter $\beta=(\rho_n-\rho_p)/(\rho_n +
\rho_p)$ for several temperatures ranging from $T = 0$ up to $T = 8$ MeV and for two
different densities. It can be observed from Fig.~\ref{fig:ratio}  that the ratio $\delta
\rho^{-}_p/\delta \rho^{-}_n$ is positive in all considered cases. This result implies
that the spinodal instability is dominated by mechanical or isoscalar density
fluctuations, which is in agreement with previous
investigations~\cite{margueron:2003,chomaz:2004,margueron:2007,vidana:2008}. Since 
$\delta \rho^{-}_p/\delta \rho^{-}_n$ is larger than the ratio  $\rho_p/\rho_n$ presented
by the black dotted line in Fig.~\ref{fig:ratio}, the dense liquid phase  of the system
is driven towards  a more isospin symmetric composition and the gas phase will be more
isospin asymmetric. These effects are less strong for higher temperatures. The liquid
phase is in general also less strongly driven towards  a isospin symmetric composition,
when the density $\rho_B$ or the asymmetry parameter $\beta$ increase.

\begin{figure}[!t]
\begin{center}
\includegraphics[width=0.7\textwidth]{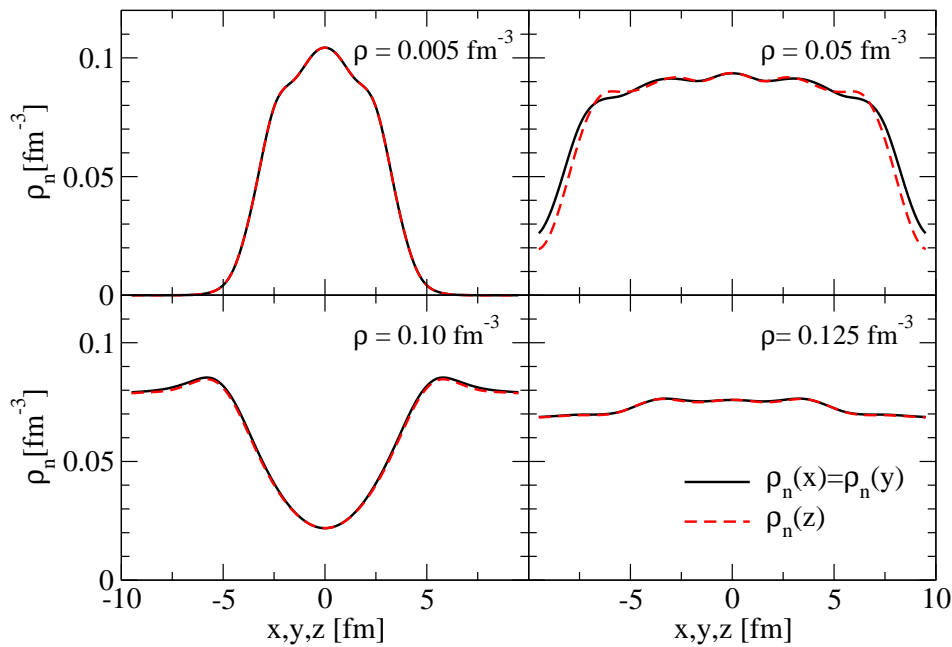} 
\caption{(Color online) Neutron density distributions at T = 0 MeV resulting from box calculations for isospin symmetric nuclear matter as a function of the Cartesian coordinates x,y, and z. The panels refer to densities of $\rho$ = 0.005 fm$^{-3}$, 0.05 fm$^{-3}$, 0.10 fm$^{-3}$, and 0.125 fm$^{-3}$. } 
\label{fig:inhombox}
\end{center}
\end{figure}
\begin{figure}[!b]
\begin{center}
\includegraphics[width=0.36\textwidth]{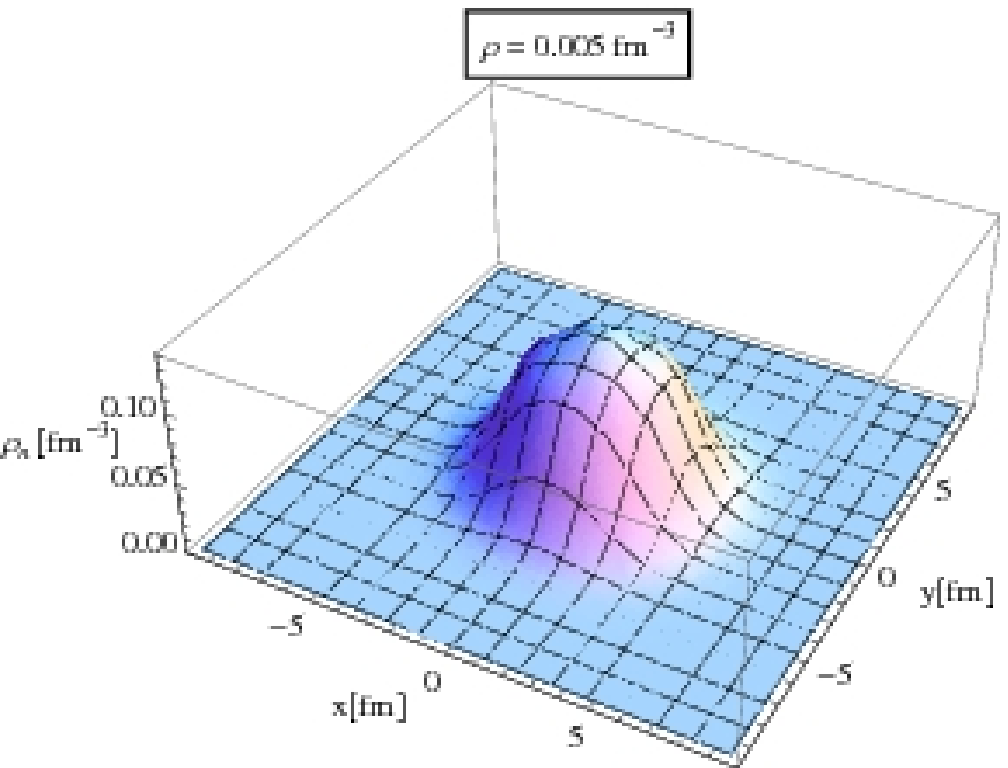} \quad \quad
\includegraphics[width=0.36\textwidth]{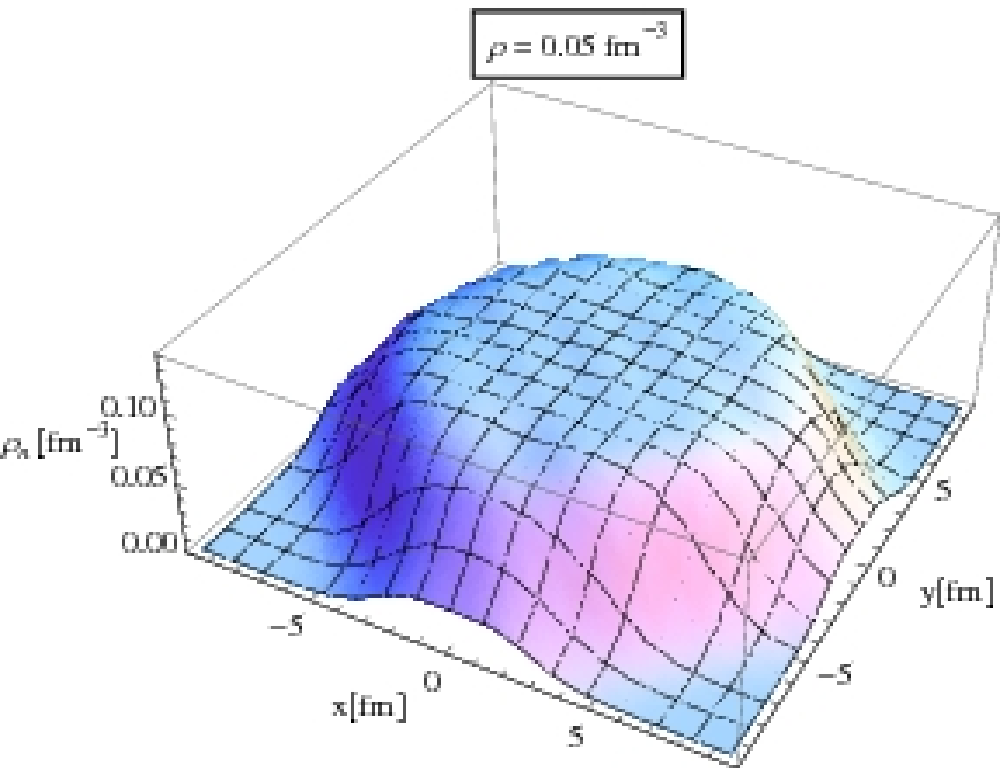} \\
\includegraphics[width=0.36\textwidth]{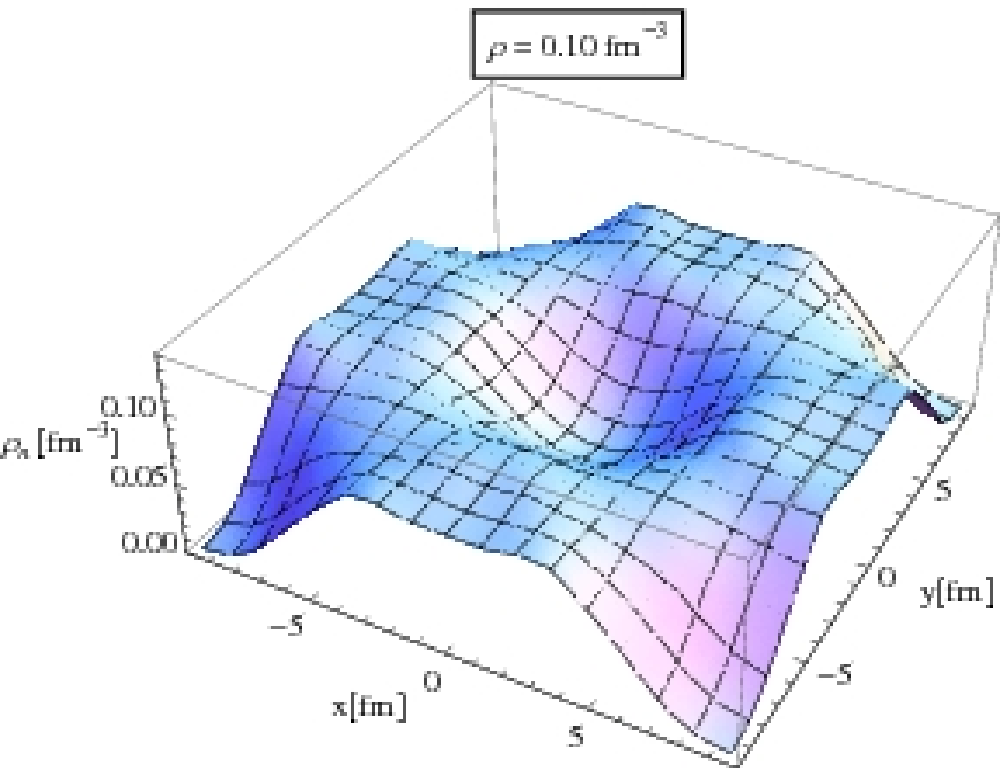} \quad \quad
\includegraphics[width=0.36\textwidth]{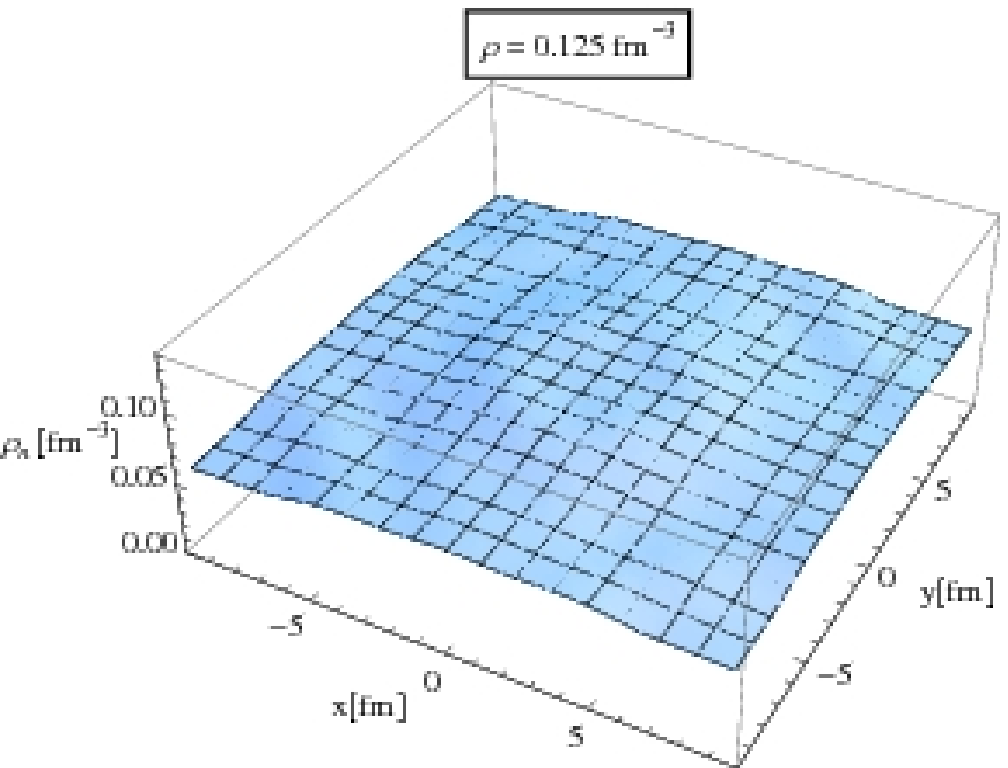}  
\caption{(Color online) Neutron density distribution at T = 0 MeV
in the z = 0 plane resulting from box calculations for isospin symmetric nuclear matter. The panels refer to densities of $\rho$ = 0.005 fm$^{-3}$, 0.05 fm$^{-3}$, 0.10 fm$^{-3}$, and 0.125 fm$^{-3}$.} 
\label{fig:inhombox2}
\end{center}
\end{figure}
Up to this point we have determined the spinodal zone from the evaluation of the
curvature matrix $[\mathcal{F}_{ij}]$ in (\ref{eq:curvature}) for the free energy of the
homogeneous system of nuclear matter. Having identified the regions of instability we
will now try to verify these instabilities and determine the density profile in this
area.

For that purpose we consider the very same DDRH  and perform variational calculations 
in a cubic Wigner-Seitz (WS) cell. The size and shape of the WS cell should be determined
in a variational manner, however, in a kind of explorative study we have chosen     
the length of the WS cell  to be 20 fm in each
direction with the origin of the coordinate system in the center of the box. Also we have
restricted the study to isospin symmetric systems. In order to be consistent 
with our infinite nuclear matter calculations, the Coulomb energy is 
not included in the WS cell calculations. Therefore the proton density distributions are
identical to the corresponding neutron ones displayed in Fig.~\ref{fig:inhombox}. 

The transition from isolated
nuclei to a phase of homogeneous baryon matter can be observed  in the various panels
Fig.~\ref{fig:inhombox} for zero temperature.  The left top panel shows the density distribution at a density
of $\rho$ = 0.005 fm$^{-3}$, i.e. only 20 protons and 20 neutrons ($^{40}$Ca) are present in the
cubic WS cell. At these low densities, isolated nuclei will be present. The
density profiles are identical in all three Cartesian directions, which corresponds to
crystal of spherical nuclei located at the centers of the WS cell. 

However, at higher
densities this is not the case anymore. At a density of $\rho$ = 0.05 fm$^{-3}$, the
density profile in $z$-direction deviates a little from that in the $x$ and $y$
direction. Furthermore, one has a high-density region at the center of the cubic WS cell
and a low-density region along the connection lines to the nearest neighbor at the 
boundary of the box. 

However, the density distribution is not as simple as suggested by
Fig.~\ref{fig:inhombox}. This can be seen from the complete density distribution in the z
= 0 plane as displayed in Fig.~\ref{fig:inhombox2}. Apart from the low density regions at
the boundary of the box along the connection lines between the centers of the WS cells,
one has also regions in which the density is almost zero. Therefore, the geometrical
structure at this density can be described by large quasi-nuclei located in the centers
of the WS cell, which are connected to each other by low-density bridges.

At even higher densities, i.e. around $\rho$ = 0.10 fm$^{-3}$, the opposite density
distribution can be observed, i.e isolated low-density regions completely surrounded by
high-density matter. At a density of $\rho$ = 0.125 fm$^{-3}$, so just a little bit below
the critical density ($\rho_c$ = 0.127 fm$^{-3}$) predicted from the study of the 
curvature matrix in homogeneous one obtains a grid
structure with small fluctuations, which resembles homogeneous nuclear matter.

Therefore, within the numerical accuracy we regard this result to be in agreement 
with the result of the infinite nuclear matter calculation in Fig.~\ref{fig:spinodal}. 

\begin{figure}[!t]
\begin{center}
\includegraphics[width=0.355\textwidth]{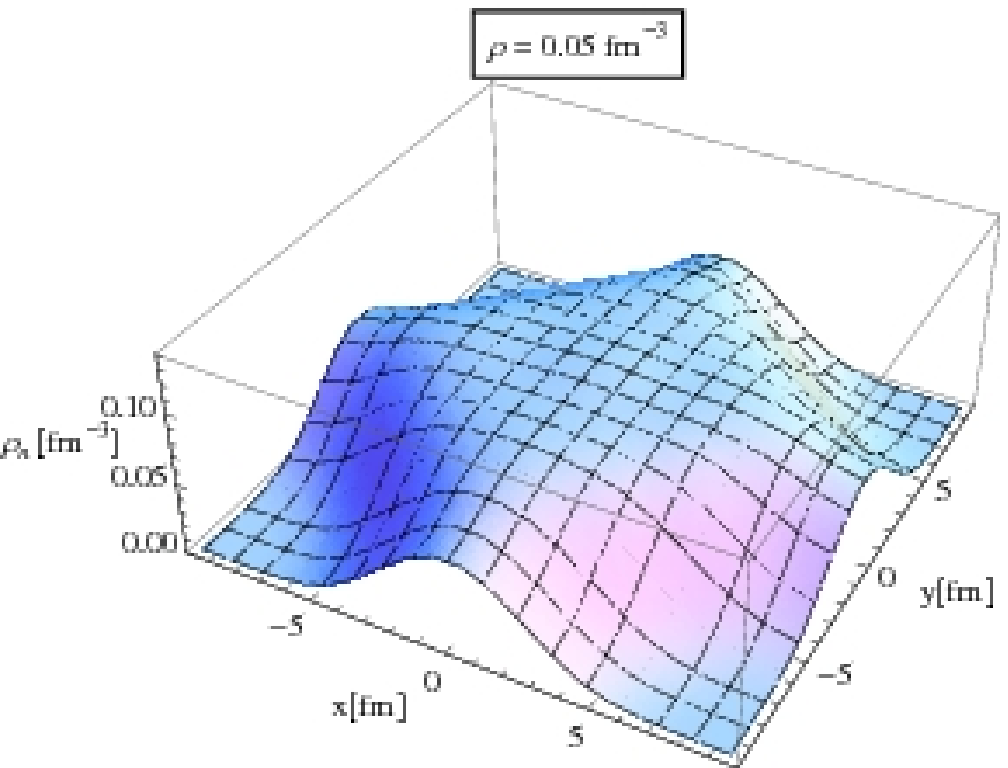} \\
\includegraphics[width=0.355\textwidth]{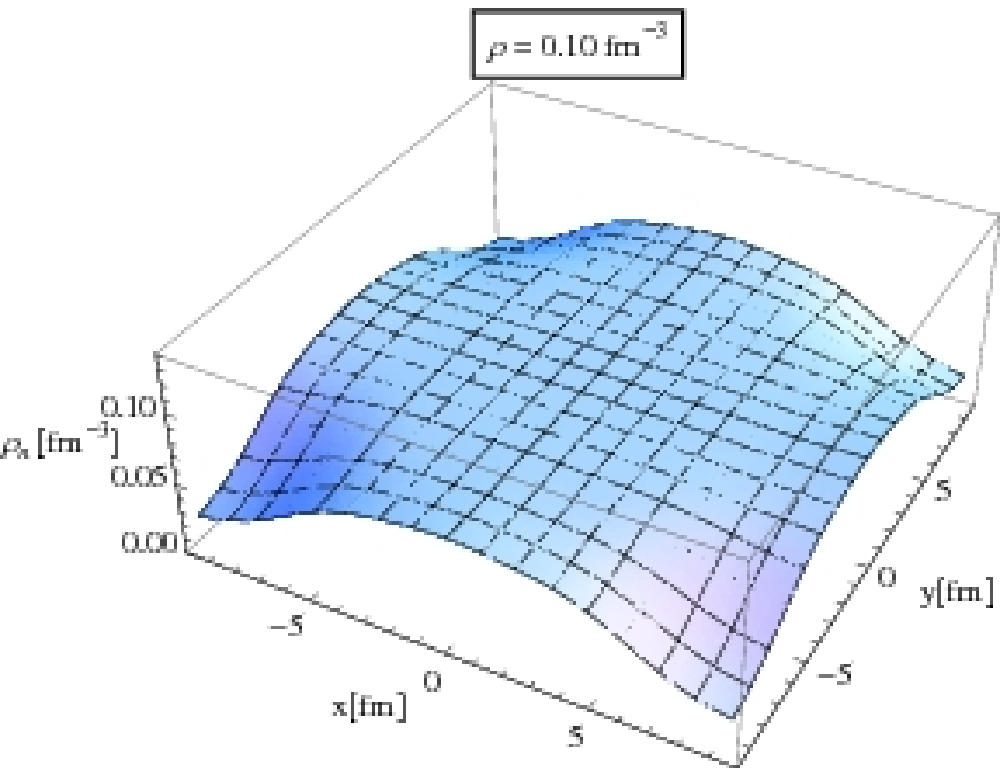} 
\includegraphics[width=0.355\textwidth]{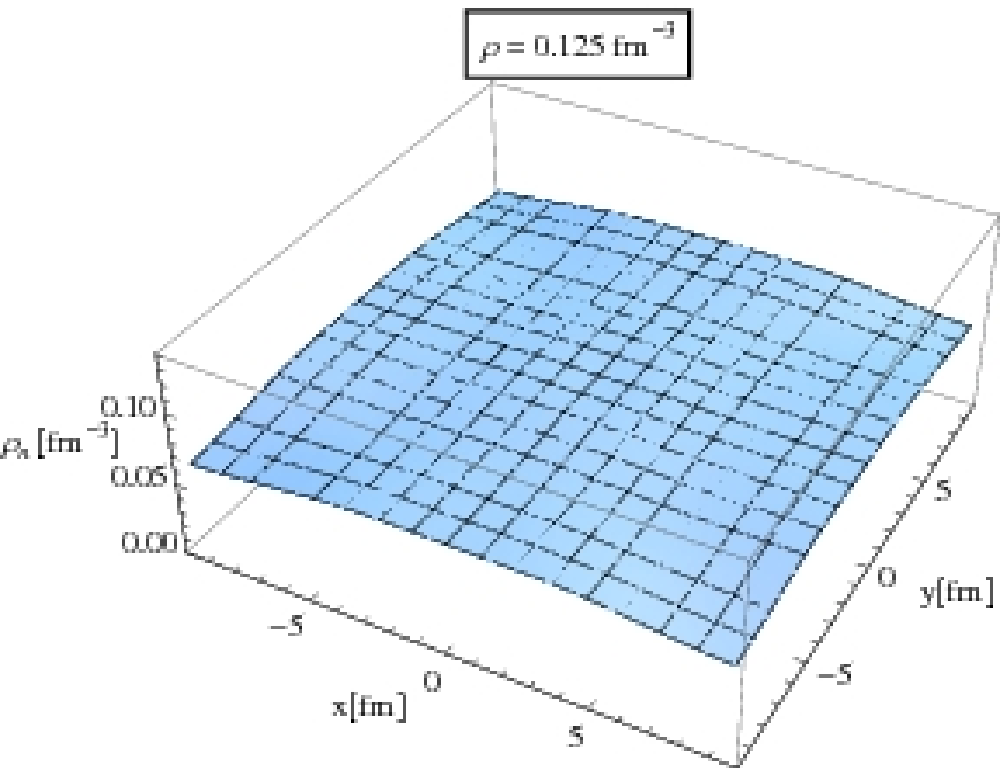}  
\caption{Neutron density distribution at T = 8 MeV
in the z = 0 plane resulting from box calculations for isospin symmetric nuclear matter. The panels refer to densities of $\rho$ = 0.05 fm$^{-3}$, 0.10 fm$^{-3}$, and 0.125 fm$^{-3}$.} 
\label{fig:inhomtembox}
\end{center}
\end{figure}
Performing DDRH calculations in the same kind of WS cell at the finite temperature T = 8
MeV, we obtain the neutron density distributions displayed in Fig.~\ref{fig:inhomtembox}.
It is clear that finite temperature effects tend to dissolve  the geometrical structures
observed for T = 0 MeV. The structures at T = 8 MeV are less pronounced than those for T
= 0 MeV in  Fig.~\ref{fig:inhombox2} and the transition density is lower at T = 8 MeV.
These results are in agreement with the results presented in Fig.~\ref{fig:spinodal} of
the infinite nuclear matter calculations.

\section{Summary and Conclusion}
\label{sec:S&C}
In the present work, we construct a Density Dependent Relativistic Mean Field model
(DDRH) to parameterize the results of microscopic Dirac Brueckner Hartree Fock (DBHF)
calculations~\cite{vandalen:2007} using the realistic Bonn A
potential~\cite{machleidt:1987}. Special attention is devoted to reproduce the 
DBHF self-energy in particular at densities below  the saturation density.  
The equation of state at zero and
finite temperature can then be determined in the framework of this DDRH theory for
homogeneous matter as well as inhomogeneous structures. The spinodal
instabilities of homogeneous matter including their temperature effects are 
investigated  within this model. 

Our investigations confirm the main results of earlier studies  
based on phenomenological~\cite{margueron:2003,chomaz:2004}
as well as realistic models of the NN interaction~\cite{vidana:2008}, that the spinodal 
instability is dominated by
mechanical or isoscalar density fluctuations which lead to a liquid-gas phase separation
with restoration of the isospin symmetry in the liquid phase. However, this restoration
is less strong for higher temperatures. 

Furthermore, the DDRH model predicts some
details, which are different from earlier studies: A large critical density at
zero temperature, a smaller critical temperature $T_c$ and details in the shape
of the spinodal region.
At zero temperature the spinodal region reaches densities above 0.12 fm$^{-3}$.
This is different from the predictions of phenomenological 
calculations~\cite{margueron:2003,chomaz:2004,avancini:2004,avancini:2006,ducoin:2006,ducoin:2007a,ducoin:2007b,liu:2002,providencia:2006,pais:2009}, but in line with the microscopic studies of Vidana and Polls~\cite{vidana:2008}. It can to some extent be attributed to the differences in the predicted saturation point of nuclear
matter. The present DDRH approach yields a smaller critical temperature
($T_c$ less than 11 MeV) and higher densities for the spinodal region at temperatures
just below $T_c$.  The predicted spinodal region obtained from our calculation is
more strongly concentrated around isospin symmetric nuclear matter.  

The studies of the spinodal regions which are based on the curvature of the free energy 
of homogeneous matter are supplemented by variational calculations allowing for density
fluctuations explicitly. Calculations employing the same DDRH
model are performed in a cubic Wigner Seitz cell by minimizing the total energy to obtain density
distributions in isospin symmetric nuclear matter. At a density of 0.125 fm$^{-3}$, a
geometrical structure resembling homogeneous nuclear matter can be found. At lower
densities, non-homogeneous nuclear matter structures occur. Finite temperature effects
tend to dissolve  the geometrical structures observed, i.e. the structures at finite
temperature  are less pronounced and the transition density to homogeneous nuclear matter
is lower.

\begin{acknowledgments}
This work has been supported by
the Deutsche Forschungsgemeinschaft (DFG) under contract no. Mu 705/7-1.
\end{acknowledgments}



\end{document}